\documentclass[useAMS,usenatbib]{mn2e_fake}
\usepackage{amsmath}
\usepackage{amssymb}
\usepackage{times} 
\usepackage{graphicx}

\usepackage{hyperref}
\usepackage{ulem}
\usepackage{relsize}

\def\mref#1{(\ref{#1})}

\DeclareMathAlphabet\EuRoman{U}{eur}{m}{n}
\SetMathAlphabet\EuRoman{bold}{U}{eur}{b}{n}
\newcommand{\eurom}{\EuRoman}
\newcommand{\yt}{{\larger\texttt{yt}}}
\renewcommand{\Im}{\textrm{Im}}

\title[SI in the quasi-global protoplanetary discs]
{Streaming instability in the quasi-global protoplanetary discs}
\author[K. Kowalik et al.]{
  K. Kowalik$^{1}$\thanks{E-mail: Kacper.Kowalik@astri.umk.pl},
  M. Hanasz$^{1}$\thanks{E-mail: Michal.Hanasz@astri.umk.pl},
  D. W\'olta\'nski$^{1}$ and
  A. Gawryszczak$^{2}$\\
  $^{1}$Centre for Astronomy, Faculty of Physics, Astronomy and Informatics,
  Nicolaus Copernicus University, Grudziadzka 5, 87-100 Toru\'n, Poland\\
  $^{2}$Pozna\'n Supercomputing and Networking Centre, Noskowskiego 10,
  61-704 Pozna\'n, Poland
  }

\begin{document}

\date{Accepted 2013 June 17. Received 2013 June 13; in original form 2013 May 15}

\pagerange{\pageref{firstpage}--\pageref{lastpage}} \pubyear{2013}

\maketitle

\label{firstpage}

\begin{abstract}
We investigate streaming instability using two-fluid approximation (neutral gas
and dust) in a quasi-global, unstratified protoplanetary disc, with the help of
{\sc PIERNIK} code. We compare amplification rate of the eigen-mode in numerical
simulations, with the corresponding growth resulting from the linear stability
analysis of full system of Euler's equation including aerodynamic drag.

Following~\cite{YG05} we show that (1) rapid dust clumping occurs due to the
difference in azimuthal velocities of gas and dust, coupled by the drag force,
(2) initial density perturbations are amplified by several orders of magnitude.
We demonstrate that the multi-fluid extension of the simple and efficient
Relaxing TVD scheme, implemented in PIERNIK, leads to results, which are
compatible with those obtained with other methods.
\end{abstract}

\begin{keywords}
hydrodynamics -- instabilities -- planets and sattelites: formation.
\end{keywords}

\section{Introduction}

The formation of planets is a complex process that requires sub-$\mu$m dust
grains to grow over a dozen orders of magnitude in size. The smallest particles,
although they are highly-coupled to gas, can slowly drift in both radial and
vertical direction, and collide with each other. That physical collisions with
low relative velocities lead to creation of larger agglomerates~\citep{BW08}.
On the other side of the scale, $10^4$~m bodies are large enough to be almost
completely decoupled from gas dynamics and be mostly influenced by gravitational
interactions~\citep{KKI06} 

\par The most problematic is the intermediate step of growing dust from cm
grains to km sized bodies, due to existence of several processes that counteract
the possible growth or pose significant time constrains. The latter is mainly
caused by a fast radial drift of the bodies that are marginally coupled to the
gas, i.e. characteristic scale of their aerodynamic drag is of the order of
their orbital time~\citep{W77}. Moreover, the characteristic relative velocities
for the bodies of sizes ranging from $1\textrm{ cm}$ to $ 1\textrm{ m}$ are of
the order of $1\div10\textrm{ m s}^{-1}$ which, during collisions, result in
fragmentation and bouncing rather than sticking~\citep{Z10}.

\par One of the possible scenarios  is a fast enhancement
of dust density via sedimentation to the midplane due to vertical component of
host star gravity. Then further fragmentation and collapse of dense dust layer
is caused by a self-gravity~\citep{GW73}. Though the sedimentation itself leads
to the rise of Kelvin-Helmholz instability~\citep{JHK06} (KHI) which prevents
forming an infinitesimally thin layer in the midplane, recent studies show that
it is not the case for massive discs with higher than solar
metallicities~\citep{L10}.

\par However, a significant flaw behind presented reasoning is a complete
negligence of disc's global turbulence, which in turn is the only mechanism that
would allow to explain observed accretion rates assuming $\alpha$-disc
model~\citep{SS73}. Even though, the most plausible mechanism responsible for
such turbulence, i.e. magnetorotational instability~\citep{BH98}, allows for
formation of local, low-level of turbulence areas due to insufficient ionization
of gas, there are still other instabilities that stir the fluids~\citep{LP10}.

\par Despite of this unfavorable circumstances there is a process that commences
to dominate dust evolution when ratio of concentration of dust particles to gas
density approaches unity. This mechanism was first presented by~\cite{YG05}
(hereafter YG05) and named the {\it streaming instability}. It appears that
combination of dust trapping in gas pressure maxima and mutual interaction
leading to dust dragging the gas, therefore enhancing the maxima even further,
results in significant dust pile-up~\citep{J11}. Even without the presence of
self-gravity, dust concentration may be risen up to the three orders of
magnitude~\citep{JY07} (hereafter JY07) which could possibly lead to
gravitationally bound objects~\citep{J07}. Recent numerical studies of
streaming instability concentrate on various physical aspects that may influence
its evolution, that includes: the influence of wide range of dust
species~\citep{BS10a}, effects of global pressure gradients~\citep{BS10b},
stratification of discs~\citep{T12}. However, all known to authors works limit
themselves to local disc approximation. 

The main goal of this paper is to investigate the streaming instability in a
more realistic circumstances of radially extended discs. The essential points of
interest are:
(1) resolution studies of the streaming instability in a 2D radially extended
domain,  to see what is the minimum  grid resolution necessary to ensure growth
rates of  simulated instability modes consistent with the results of linear
stability analysis, (2) investigation of the streaming instability a in 3D
radially extended domain, in an optimal  grid resolution,  deduced from
resolution studies in 2D.

The plan of the paper is as follows: In Section~\ref{sec:eqs} we set up the
basic assumptions and equations relevant for the present paper. In
Section~\ref{sec:lsa} we perform linear analysis of streaming instability in
protoplanetary discs. In Section~\ref{sec:setup} we describe specific numerical
algorithm that were used in the performed simulations along with the initial
conditions and the basic simulation parameters. Section~\ref{sec:results} shows
the obtained results and their detailed comparison with linear stability
analysis.  Finally in Section~\ref{sec:conclusions} we discuss the outcome of
our experiments and give short outlook of the future work.

\section{Basic assumptions and equations}
\label{sec:eqs}

We investigate global dynamics of two interacting fluids -- gas and dust -- in
the protoplanetary disc. We assume that neutral gas obeys the isothermal
equation of state, whereas dust is treated as a pressureless fluid.

\begin{align}
\partial_t \rho_g &+ \nabla\cdot\left(\rho_g\mathbf{u}\right) = 0,\\
\partial_t \rho_d &+ \nabla\cdot\left(\rho_d\mathbf{w}\right) = 0,\\
\partial_t \left(\rho_g\mathbf{u}\right) &+
   \nabla\cdot(\mathbf{u}\otimes(\rho_g\mathbf{u})+P) \notag\\
 &= -\rho_g\left(\nabla\Phi +
\frac{\rho_d}{\tau_f\rho_g}(\mathbf{u}-\mathbf{w})\right),\label{eq3}\\
\partial_t \left(\rho_d\mathbf{w}\right) &+
\nabla\cdot(\mathbf{w}\otimes(\rho_d\mathbf{w})) \notag\\
 &= -\rho_d\left(\nabla\Phi + \frac{1}{\tau_f}(\mathbf{w}-\mathbf{u})\right)
\label{eq4}.
\end{align}

\noindent where $\rho_g$, $\rho_d$ are densities of gas and dust respectively,
$\mathbf{u}$, $\mathbf{w}$ corresponding velocities, $P$ gas pressure, $\tau_f$
is a friction time and $\Phi$ is gravitational potential. Our aim is to isolate
the streaming instability from the other possible processes acting in a radially
extended protoplanetary discs. In order to obtain a clear diagnostics of the
excited instability we neglect the vertical component of gravity arising from the
point mass source in the center of the reference frame, which would lead to dust
sedimentation and eventually to the onset of KHI~\citep{JHK06}.

We shall simulate disc starting from the relatively large radius (2~AU).  We
assume, following~\cite{CD93} that transition to Stokes regime occurs for
particle radius $a = 9/4\lambda_g$ where  $\lambda_g = 4.2\times 10^4\textrm{
cm} (10^{-14}\textrm{ g cm}^{-3}/\rho_g) \approx (R/1 \textrm{AU})^{2.75}$~cm is
the mean free path of the gas molecules~\citep{W77,BT09},  and $R$ is the radial
distance to the disc centre.  Under these assumptions the Epstein's regime
applies in the dominating part our domain, even for the largest simulated grain
sizes.  Therefore, to compute the friction time, we use the Epstein's law
\begin{equation}
   \tau_f = \frac{\rho_\bullet a} 
      {\rho_g \sqrt{c_s^2 + |\mathbf{u} - \mathbf{w}|^2 }}
   \label{eq:tauf} 
\end{equation}
where $c_s$ is the gas sound speed, and $\rho_\bullet = 1.6\textrm{ g cm}^{-3}$
is the density of the solid material.
While the term $|\mathbf{u}-\mathbf{w}|^2$ does not play any significant
role during the evolution of SI, i.e. its value never exceeds 1 per cent of
$c_s^2$ we have decided not to neglect it in our simulations for the
completeness sake.

\begin{figure*}
   \centering
   \begin{tabular}{@{}cc@{}}
      \includegraphics[width=0.49\linewidth]{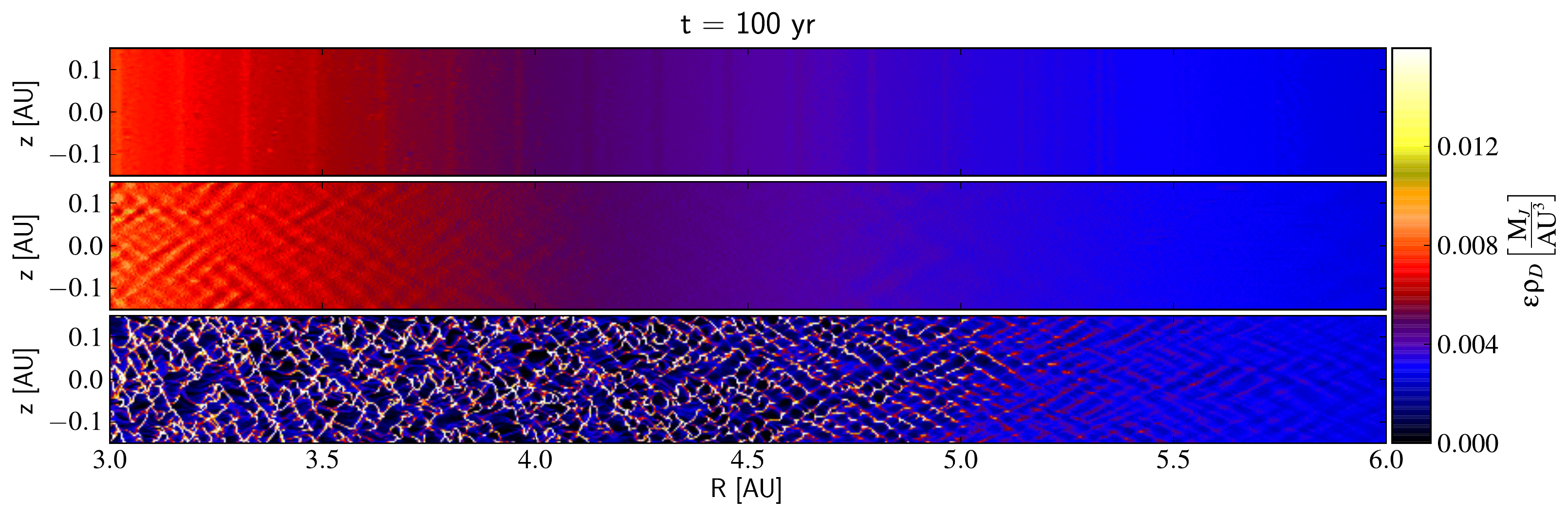} & 
      \includegraphics[width=0.49\linewidth]{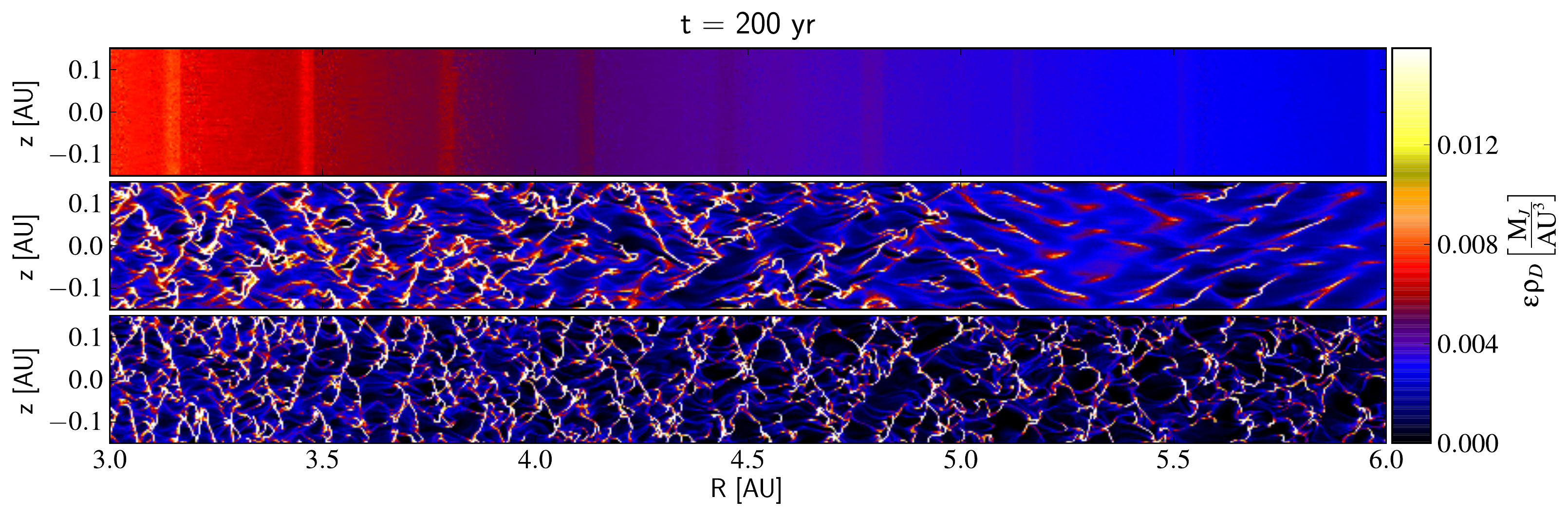} \\
      \includegraphics[width=0.49\linewidth]{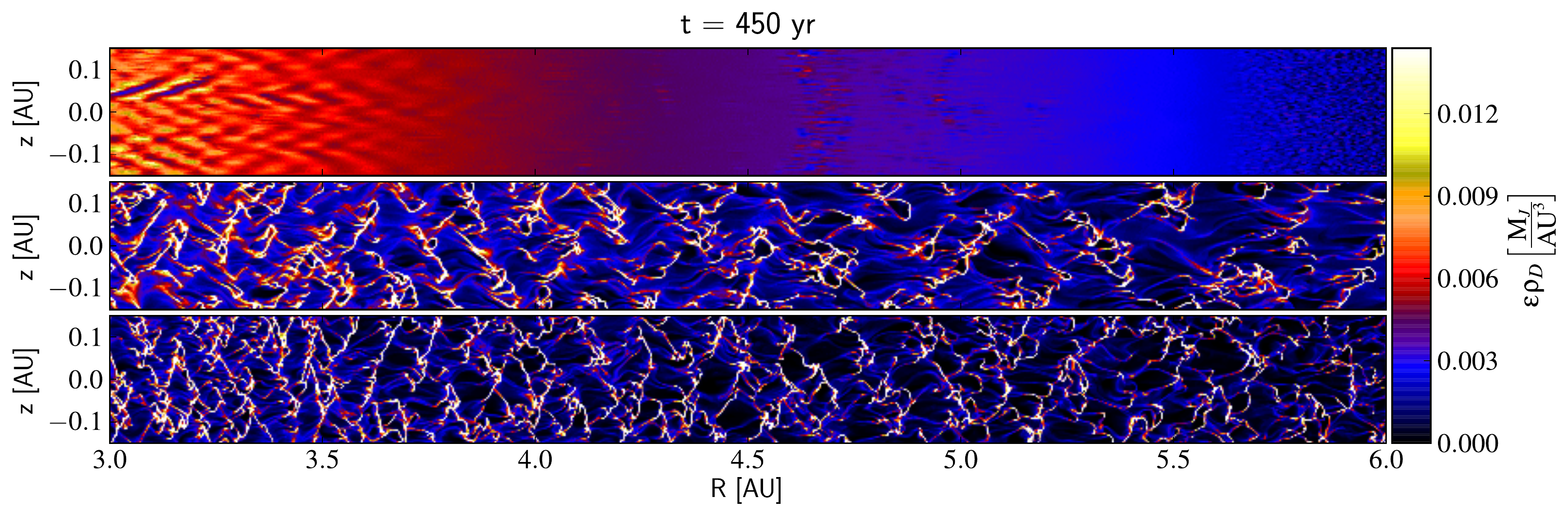} &
      \includegraphics[width=0.49\linewidth]{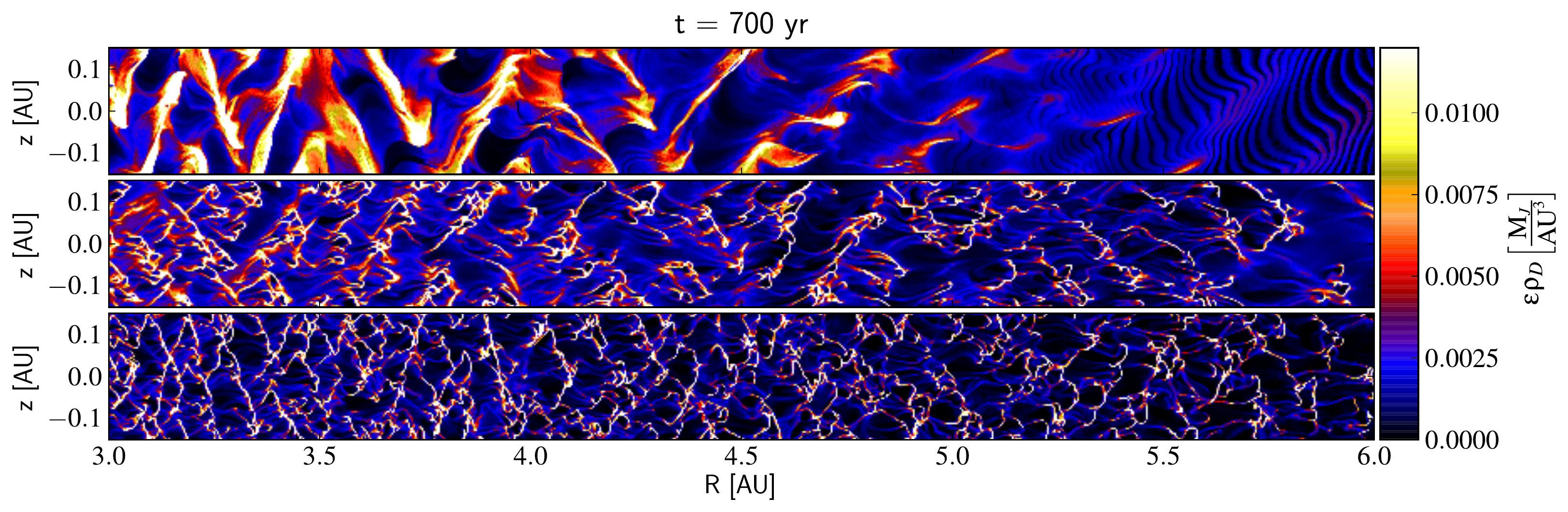}
   \end{tabular}
   \caption{Dust density snapshots for runs with $50$~cm grain sizes at $100$,
   $200$, $450$ and $700$~yr for upper left, upper right, bottom left and bottom
   right panel respectively. Each of the panels is divided into three subpanels
   for initial $\epsilon = 0.2, 1, 2.0$ on upper (BAh), middle (BB) and lower
   (BC) subpanel respectively. Animation of the run BBh can be found 
   \href{http://youtu.be/EfQU8ixHFbU}{here}.}
   \label{fig1}
\end{figure*}
\section{Linear stability analysis}
\label{sec:lsa}
One of the aims of this paper is to validate numerical results related to the
growth of streaming instability unstable modes via the comparison to the growth
rate and mode geometry resulting from linear stability analysis. Therefore,
following the previous authors~\citep{YG05}, in this section, we review the main
steps of the linear analysis of the streaming instability in protoplanetary
discs.

The linear stability analysis relies on three essential elements: (1) finding an
equilibrium state of the system, (2) perturbing the system with small amplitude
(linear) perturbations, and (3) deriving the eigenmodes and their growth rates. 
Unfortunately there is no strictly stationary solution to global Keplerian disc
consisting of two fluids coupled via the drag force. 
The reason is the dust migration, implying variations of the radial disc profile.
We note however, that the characteristic timescale of radial
migration of meter-size dust agglomerates is of the order of 100 years or
longer, therefore we consider the variations of the disc profile as a
slow process, as compared to typical growth times of  streaming instability.

Another difficulty results from the fact that the wide range of the radial
coordinate in our numerical models implies radial variations of the equilibrium
state the physical parameters. In these circumstances it would be appropriate to
perform global stability analysis by means of solving the two-point boundary
value problem (see e.g. \cite{PHM04, KH06}), however the latter approach is
significantly more complex than the local stability analysis. Therefore, we
consider the local stability analysis as a first approximation to the full
analysis of streaming-instability in protoplanetary discs. The modes derived in
the framework of local linear stability analysis will serve us as a reference
solution for validation of the modes excited in the global numerical
simulations.

The most suitable local treatment of streaming instability is provided by the
"shearing-sheet" approximation~\citep{HGB95}, which is achieved by placing
Cartesian coordinate frame on an orbit of radius $R$, that corotates with
Keplerian frequency $\Omega$.  We assume that $x$ axis points radially outward,
and $y$ corresponds to the azimuthal direction, whereas $z$ is a vertical axis.
Following \cite*{YJ07} we write down the continuity and Euler equations for both
the gas and dust components
\begin{align}
\partial_t \rho_g &+ \mathbf{u}\cdot\nabla\rho_g - \frac{3}{2}\Omega x\partial_y\rho_g 
 = -\rho_g\nabla\cdot\mathbf{u},\label{eqc1}\\
\partial_t \rho_d &+ \mathbf{w}\cdot\nabla\rho_d - \frac{3}{2}\Omega x\partial_y\rho_d 
 = -\rho_d\nabla\cdot\mathbf{w},\label{eqc2}\\
\partial_t \mathbf{u} &+ \left(\mathbf{u}\cdot\nabla\right)\mathbf{u} 
 - \frac{3}{2}\Omega x\partial_y\mathbf{u} 
 = 2\Omega u_y \hat{\mathbf{x}} -\frac{1}{2}\Omega u_x \hat{\mathbf{y}} \notag\\
 &- \frac{\epsilon}{\tau_f}(\mathbf{u}-\mathbf{w}) -c_s^2\nabla\ln\rho_g 
 +2\eta\Omega^2 R \hat{\mathbf{x}},\label{eqm1}\\
\partial_t \mathbf{w} &+ \left(\mathbf{w}\cdot\nabla\right)\mathbf{w} 
 - \frac{3}{2}\Omega x\partial_y\mathbf{w}
 = 2\Omega w_y \hat{\mathbf{x}} -\frac{1}{2}\Omega w_x \hat{\mathbf{y}} \notag\\
 &- \frac{1}{\tau_f}(\mathbf{w}-\mathbf{u}), \label{eqm2}
\end{align}
where transport terms, such as $(3/2)\Omega x$ on the left-hand side of
equations arise due to fact that we measure all velocities relative to linear,
Keplerian shear flow in the rotating frame $\mathbf{v}_0 = -(3/2)\Omega x
\hat{\mathbf{y}}$. It is worth noting that term $-(1/2)\Omega \{u,w\}_x
\hat{\mathbf{y}}$ on the right hand side of the equations of motion
\mref{eqm1}-\mref{eqm2} is a sum of two components: $(-2\Omega \{u,w\}_x +
(3/2)\Omega \{u,w\}_x) \hat{\mathbf{y}}$ where the first one is the component of
Coriolis force and the second results from aforementioned subtraction of the
mean flow. The main difference between equations \mref{eqm1} and \mref{eqm2} is
the radial pressure gradient term influencing  the dynamics of gas only.  As
noted by YG05, it is possible, within the local approximation,  to include
consistently the disc's global pressure gradient parametrized by the
dimensionless measure of sub-Keplerian rotation

\begin{equation}
\eta \equiv - \frac{\partial_R P}{2\rho_g\Omega^2 R} \sim \frac{c_s^2}{v_K^2}.
\end{equation}

The set of equations \mref{eqc1}-\mref{eqm2} has a known equilibrium
solution~\citep{N86}

\begin{align}
\bar{\mathbf{w}} &= \left[ 
 -2\tau_s\xi, \frac{\tau_s^2\xi - 1}{1+\epsilon},
 0
\right]\eta v_K, \label{eq:w0}\\
\bar{\mathbf{u}} &= \left[ 
 2\epsilon\tau_s\xi, -\frac{1 + \epsilon\tau_s^2\xi}{1+\epsilon},
 0
\right]\eta v_K, \label{eq:u0}
\end{align}
where $\tau_s = \Omega \tau_f$ is dimensionless stopping time and
$\xi = ((1+\epsilon)^2 + \tau_s^2)^{-1}$. We linearize eq.
\mref{eqc1}-\mref{eqm2}, decomposing variables into a steady part and a
perturbation $\mathbf{q} = \bar{\mathbf{q}} + \mathbf{q}^\prime$, where
$\mathbf{q}=[\rho_d, w_x, w_y, w_z, \rho_g, u_x, u_y, u_z]$. We assume
subsequently that the perturbations are axisymmetric (independent on
$y$-coordinate in the present analysis) and can be expressed as plane waves

\begin{equation}
   \label{eq:planar}
   \mathbf{q}^\prime(x,z,t) = \tilde{\mathbf{q}}
 \exp\left[i(k_x x + k_z z -\omega t)\right]
\end{equation}

After the substitution of the linear perturbations the equations read

\begin{align}
-i(\omega- k_x\bar{w}_x)\tilde{\rho}_d &= 
 - i \bar{\rho}_d(k_x\tilde{w}_x + k_z\tilde{w}_z), \label{lin1}\\
-i(\omega- k_x\bar{u}_x)\tilde{\rho}_g &= 
 - i \bar{\rho}_g(k_x\tilde{u}_x + k_z\tilde{u}_z), \label{lin2}\\
-i(\omega- k_x\bar{u}_x)\tilde{\mathbf{u}} &= 
 2\Omega \tilde{u}_y\hat{\mathbf{x}} - \frac{1}{2}\Omega \tilde{u}_x
 \hat{\mathbf{y}}
 -\frac{\epsilon}{\tau_f}(\tilde{\mathbf{u}} - \tilde{\mathbf{w}}) \notag\\
  &-\frac{\tilde{\rho}_d}{\bar{\rho}_g\tau_f}
  (\bar{\mathbf{u}} - \bar{\mathbf{w}})
  - \frac{c_s^2}{\bar{\rho}_g}ik_x\tilde{\rho}_g, \label{lin3}\\
-i(\omega- k_x\bar{w}_x)\tilde{\mathbf{w}} &= 
 2\Omega \tilde{w}_y\hat{\mathbf{x}} - \frac{1}{2}\Omega \tilde{w}_x
 \hat{\mathbf{y}} 
 - \frac{1}{\tau_f} (\tilde{\mathbf{w}} - \tilde{\mathbf{u}}), \label{lin4}
\end{align}
where $\epsilon = \bar{\rho}_d/\bar{\rho}_g$. The set of equations
\mref{lin1}-\mref{lin4} can be expressed as

\begin{equation}
 \eurom{A}(k_x,k_z,\omega)\tilde{\mathbf{q}} = 0
 \label{eq:linset}
\end{equation}
Nontrivial solutions of the linear system \mref{eq:linset} exist if
\begin{equation}
 \det|\eurom{A}(k_x,k_z,\omega)|=0
 \label{eq:disprel}
\end{equation}
We solve the dispersion relation \mref{eq:disprel} for given values of
$(k_x,k_z)$, with respect to the complex frequency $\omega$, using the
Durand-Kerner method~\citep{D60,K66}, and subsequently obtain relation between
constant amplitudes of components of the vector
$\tilde{\mathbf{q}}$~\footnote{Numerical solvers to both dispersion relation and
eigenvector calculation along with most visualization routines are publicly
available at \url{http://github.com/Xarthisius/kowalik_2012a}}.  The growth rate
of the instability is defined as imaginary part of the complex frequency $s=\Im
(\omega)$.
\begin{figure*}
   \centering
   \begin{tabular}{@{}cc@{}}
      \includegraphics[width=0.49\linewidth]{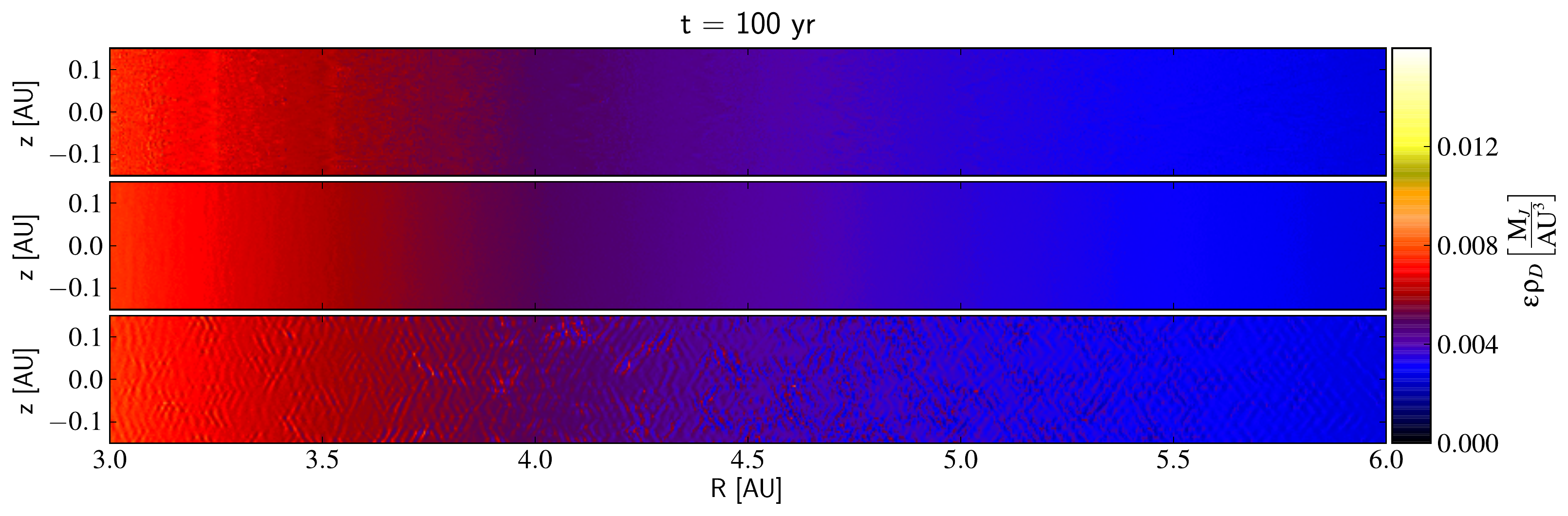} &
      \includegraphics[width=0.49\linewidth]{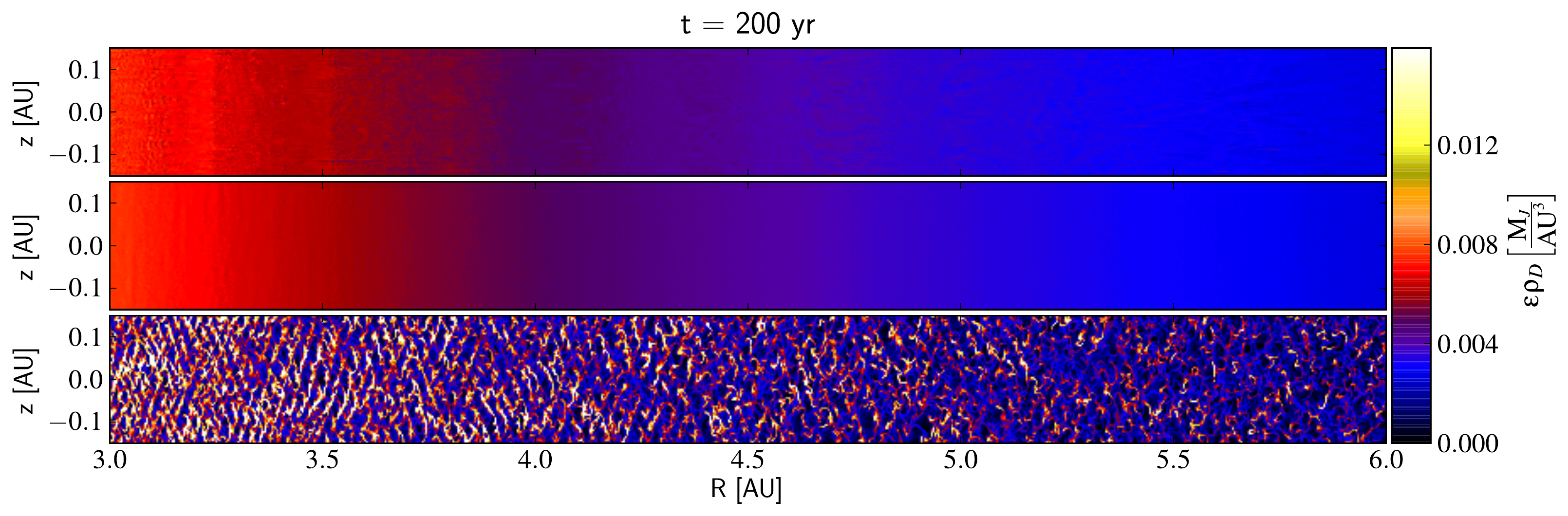} \\
      \includegraphics[width=0.49\linewidth]{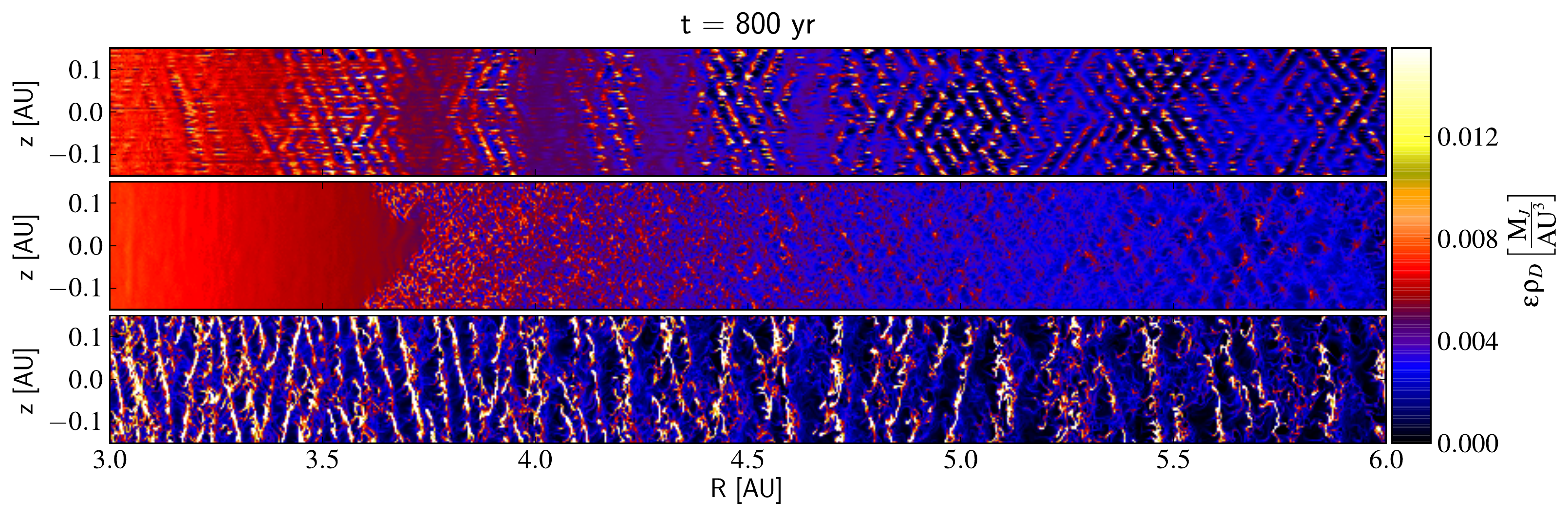} &
      \includegraphics[width=0.49\linewidth]{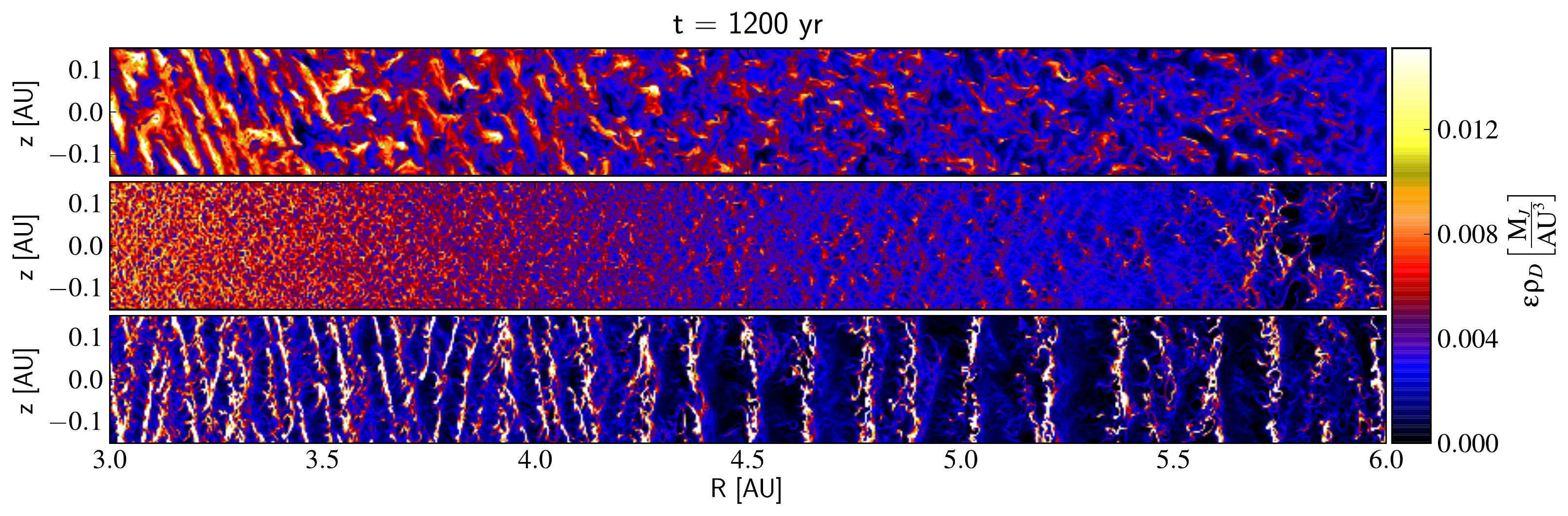}
   \end{tabular}
\caption{Dust density snapshots for runs with $10$~cm grain sizes at $100$,
   $200$, $800$ and $1200$~yr for upper left, upper right, bottom left and
   bottom right panel respectively.  Each of the panels is divided into three
   subpanels for initial $\epsilon = 0.2, 1, 2.0$ on upper (AA), middle (AB) and
   lower (AC) subpanel respectively. Animation of the run AAu can be found 
   \href{http://youtu.be/b7myS1PA_tE}{here}.}
\label{fig2}
\end{figure*}
%
\section[Simulation Setup]{Simulation setup}
\label{sec:setup}
\subsection{Algorithms}

We conduct numerical simulations with the aid of a parallel MHD code PIERNIK
using the cylindrical coordinate system. 
Following \cite{M07} and \cite{SO10} we use ''angular momentum-conserving form''
of the $\phi$-momentum equation, which with respect to Cartesian geometry
introduces only one additional source term to equations~\mref{eq3} - \mref{eq4}:
$\left((\rho_g u_\phi + P) / R\right)\mathbf{\hat{R}}$ and $(\rho_d w_\phi / R)
\mathbf{\hat{R}}$ respectively.

Both components i.e. gas and dust are treated as fluids~\citep{piernik2} which
are dynamically coupled via the friction force. In order to prevent large
timestep constraint from drag force acceleration we used the semi-implicit
scheme by~\cite{TB09}. We allow the system to relax numerically (during the
period of initial $10$~yrs of the simulation) before we ''turn on'' the
aerodynamic drag force and seed dust velocities with low--amplitude random
noise. The feedback of the linear drag force scales with the density ratio of
dust to gas $\epsilon\equiv \rho_d/\rho_g$.

\subsection{Initial disc configuration}

The initial density profile for gas roughly follows prescription of Minimal Mass
Solar Nebula~\citep{H81}
\begin{equation}
   \Sigma(R) = 1700 \left(\frac{R}{1\textrm{ AU}}\right)^{-3/2} 
   \textrm{ g cm}^{-2}.
\end{equation}
We assume isothermal equation of state and a constant temperature $T_0 = 170$~K
across the whole disc. 
We assume gravitational field from a point mass $M=1\,\textrm{M}_\odot$, and
neglect the vertical component of gravitational acceleration towards the central
mass, implying no vertical stratification of the disc. Yet, we refer to the
vertical scaleheight $H$ to estimate  volume density of gas at given radius,
relying on the hydrostatic equilibrium density distribution in the presence of
the vertical gravity of a point mass
\begin{equation}
   \rho(R,z) =  \rho(R,0) \exp\left(-\frac{z^2}{2H(R)^2}\right),
\end{equation}
where $\rho(R,0)$ is gas density at the disc midplane and $H^2 = 2 c_s^2 R^3/
GM$.
The surface gas density is given by
\begin{equation}
   \Sigma(R) = \int_{-\infty}^\infty \rho(R,z) dz,
\end{equation}
The corresponding value of midplane gas density is
\begin{equation}
   \label{eq:rho}
    \rho(R,0) = \frac{\Sigma(R) }{\int_{-\infty}^\infty
   \exp\left(-\frac{z^2}{2H(R)^2}\right) dz}.
\end{equation}
For the chosen disc temperature $T_0$ the integral on the right hand side of
\mref{eq:rho} varies from $0.4$~AU to $2.0$~AU over the range of radii $R\in
[2,6]$~AU. 
We assume for simplicity its value equal to $1$~AU. 

We assume the vertical extent of the computational domain $L_z = 0.3$~AU, and
impose periodic boundary conditions at upper and lower $z$-boundaries. The
initial condition relies on a radial force balance for the gas and dust
components independently. The gas component remains in a hydrostatic equilibrium
resulting from the radial balance of gravity, centrifugal and pressure forces,
while the pressure gradient term is absent in the equation of motion for the
dust component.  Reflecting boundary conditions are set on the inner and outer
boundaries of the computational grid to prevent mass escape from the
computational domain. 

\par To minimise unphysical wave reflections, we implement wave killing zones
close to the inner and outer radial boundaries. The inner wave killing zones
cover 0.5 AU near the inner and outer edges of the computational domain. In
these zones, we add an additional damping term, in the evolution equations of
each fluid variable

\begin{equation}
  \frac{\textrm{d}X}{\textrm{d}t} = - \frac{X-X_0}{T_d}f(R),
\end{equation}
together with
\begin{equation}
   \begin{split} 
      f(R) &= 1 - \tanh\left(\left(R - R_\textrm{in} + 1
      \right)^{f_\textrm{in}}\right)\\ &+ \max\left\{ \tanh\left(\left(R -
      R_\textrm{out} + 1\right)^{f_\textrm{out}}\right), 0\right\}, 
   \end{split}
\end{equation}
where $X_0$ is the initial value of $X$ and $T_d$ is the damping timescale. We
chose $T_d$ of the order of orbital period.  The exponents
$f_\textrm{in}=f_\textrm{out}=10$ control the width of the transition layer
between the unmodified to the damped zones. Within the damped zones, the effects
of undesired wave reflection are essentially minimised.
\subsection{Simulation parameters}
We have performed a parameter sweep for the ratio of dust to gas density
$\epsilon$ and the size of particles $a$, at two different resolutions of the
computational grid in 2D, and additionally we have realized one of the models in
3D. Our choice of these parameters follows closely that of JY07, to enable
detailed comparison of the results.

The full list of simulations together with their main parameters is presented in
Table~\ref{tab1}. In all simulations the domain has height of $0.293$~AU,
for 2D it covers radii from 2 to 7~AU, whereas BB3D extends from 2 to 12~AU and
spans $\varphi\in[0, \pi/6]$.  Following JY07 we vary $\epsilon$ from 0.2 to
2.0~in order to exhibit morphologically different outcomes of nonlinear phase of
streaming instability.  We choose particle radii to be grater than $10$~cm and
less than $50$~cm so that (1) particles fall into Epstein regime in all
simulations, (2) their size could be plausibly explained by the growth processes
e.g. collisional agglomeration.

\begin{table}
   \centering
   \begin{tabular}{cccccc}
      \hline
      Run & $N_r \times N_\varphi \times N_z$ &
      $a$~[cm] & $\epsilon$ & $T_\textrm{end}$~[yr] \\
      \hline
      BB3D &  $5120  \times 128 \times 150$  & 50  & 1.0 & 500  \\
      AA   &  $5120  \times 1   \times 300$  & 10  & 0.2 & 3000 \\
      AB   &  $5120  \times 1   \times 300$  & 10  & 1.0 & 3000 \\
      AC   &  $5120  \times 1   \times 300$  & 10  & 2.0 & 3000 \\
      BA   &  $5120  \times 1   \times 300$  & 50  & 0.2 & 3000 \\
      BB   &  $5120  \times 1   \times 300$  & 50  & 1.0 & 3000 \\
      BC   &  $5120  \times 1   \times 300$  & 50  & 2.0 & 3000 \\
      AAh  &  $10240 \times 1   \times 600$  & 10  & 0.2 & 1700 \\
      AAu  &  $20480 \times 1   \times 1200$ & 10  & 0.2 & 1800 \\
      ABh  &  $10240 \times 1   \times 600$  & 10  & 1.0 & 1400 \\
      BAh  &  $10240 \times 1   \times 600$  & 50  & 0.2 & 1730 \\
      BBh  &  $10240 \times 1   \times 600$  & 50  & 1.0 & 3000 \\
      \hline
   \end{tabular}
\caption{Simulation parameters. Columns give, from left to right, name of the
   run, span of the computational domain in AU (r,z) and azimuthal angle, grid
   resolution, particle diameter in cm, solids-to-gas ratio, and total run time
   in units of years.} 
\label{tab1} 
\end{table}

\section{Results}
\label{sec:results}
The initial stage of streaming instability evolution is similar for all cases of
$\epsilon$ or $a$ and is governed by the dominant linear modes. Detailed
analysis of linear phase is covered in Section~\ref{simulation_analysis}. The
following two sections describe different non-linear evolution of streaming
instability in quasi-global setup with reference to similar case shown by JY07.
\subsection{Marginally Coupled Boulders ($a=50\,\textrm{cm}$, $\tau_s\approx
1.2$)\label{marg_boulders}}
As noted by JY07, streaming instability is the most prominent for $\tau_s = 1$
and exhibits fast linear growth and heavy dust clumping.  Figure~\ref{fig1}
shows temporal evolution of streaming instability for different initial dust to
gas ratios $\epsilon$. In all cases (BA, BB, BC) elongated clumps are formed.
When the saturated state is reached, due to the high vertical speed of dust
overdensities, the filaments undergo occasional fragmentation and collisions
with each other, though still following "v"-shaped trajectories that emerged
during linear phase. The most prominent differences between BA, BB and BC are:
(1) the characteristic length of local clumps is decreasing with the increase of
initial $\epsilon$ and (2) their tendency to lean in radial direction: in BB
structures are almost purely diagonal, whereas in BA and BC are much more
elongated in the vertical direction.
Similarly to JY07, in all cases (BA, BB, BC) the density peaks of the dust
component in the non-linear regime settle at levels about 2 order of magnitude
higher than initial density.

\subsection{Tightly Coupled Boulders ($a=10\,\textrm{cm}$, $\tau_s\approx 0.24$)
\label{tight_boulders}}

\begin{figure} 
\includegraphics[width=0.98\linewidth]{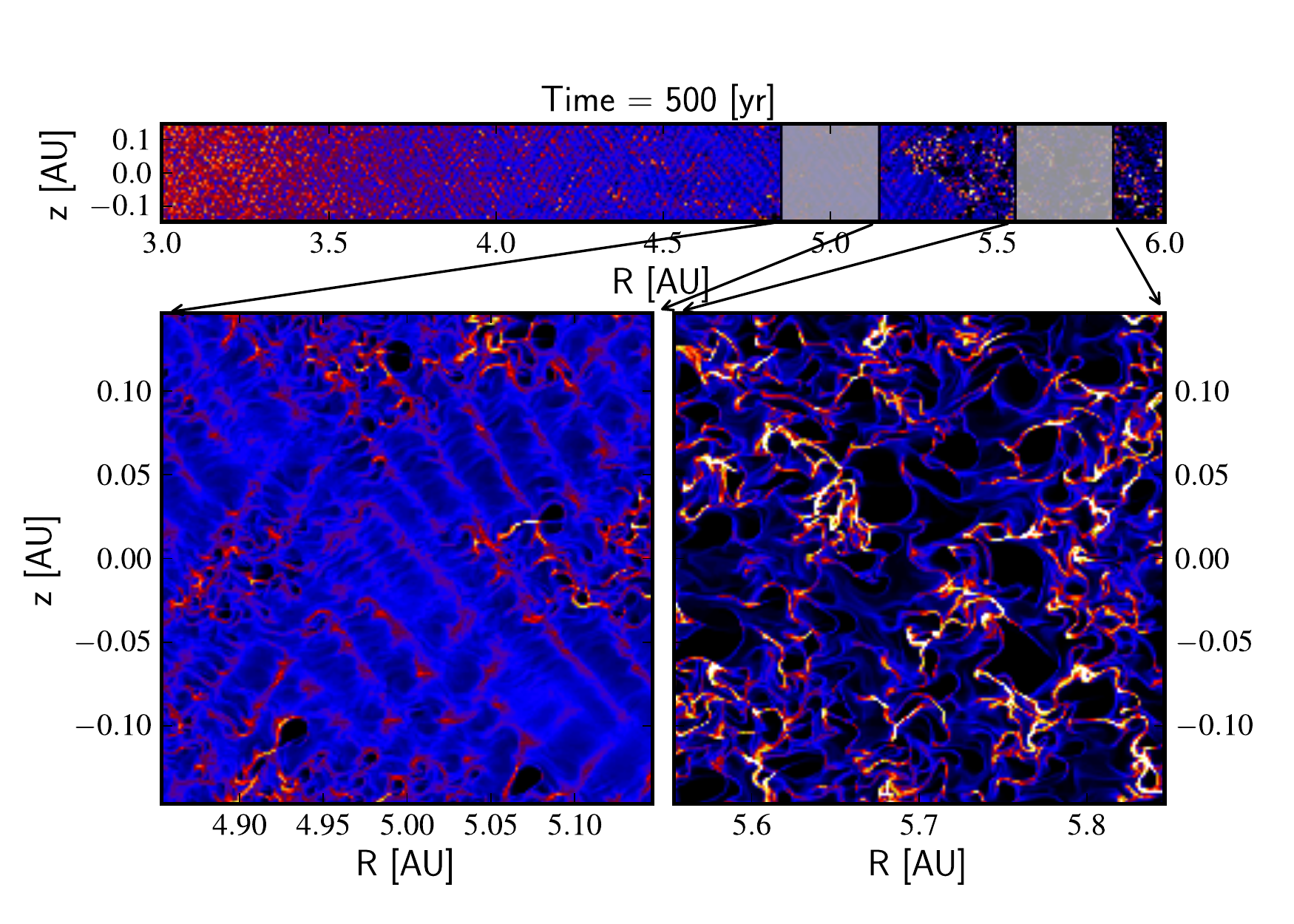}
\caption{Dust density snapshot for run ABh showing two "zoom-in" regions where
cavities emerge from overdensities created during linear phase of evolution
(left panel) and region totally overwhelmed by bursting void bubbles and vivid
dust turbulence (right panel). Animation of the run ABh can be found 
\href{http://youtu.be/NoA5-TiQabQ}{here}.}
\label{fig3}
\end{figure}

Since the fluid approach, contrary to the particle description, is much less
susceptible to effects of Poisson noise we were able to follow linear phase for
run AB and ABh for much longer than JY07, though in the end we also observe
sudden increase in growth rate of the instability due to spontaneous cavitation
(initial phase is caught in the middle subpanel of bottom right panel of
Fig.~\ref{fig2}).
The bubbles of void start to spawn at outer radii and greatly increase the dust
density and velocity on their edges (see Fig.~\ref{fig3}). After just
few orbital periods the accelerated instability expands to whole domain and
dominate the nonlinear evolution. Afterwards the quasi-stationary state of
vivid turbulence is achieved. 
In both cases (AB, AC), the non-linear evolution leads to the enhancement
of the peak densities of dust component over one order of magnitude, which is in
close agreement with results of JY07 (see Fig.~8 in their work).

\par In the dust dominated case (AC) our results closely follow those of JY07 up
to the point of saturation, where we achieve highly turbulent flow (lower
subpanel of upper right panel of Fig.~\ref{fig2}) and about 20 fold growth
in dust density. However, we note that secular evolution of the physical system
leads to formation of elongated dust overdensities and further growth up to 2
orders of magnitude as seen in runs with other parameters (see lower subpanel of
bottom right panel of Fig.~\ref{fig2}).

\par
\begin{figure}
   \includegraphics[width=0.98\linewidth]{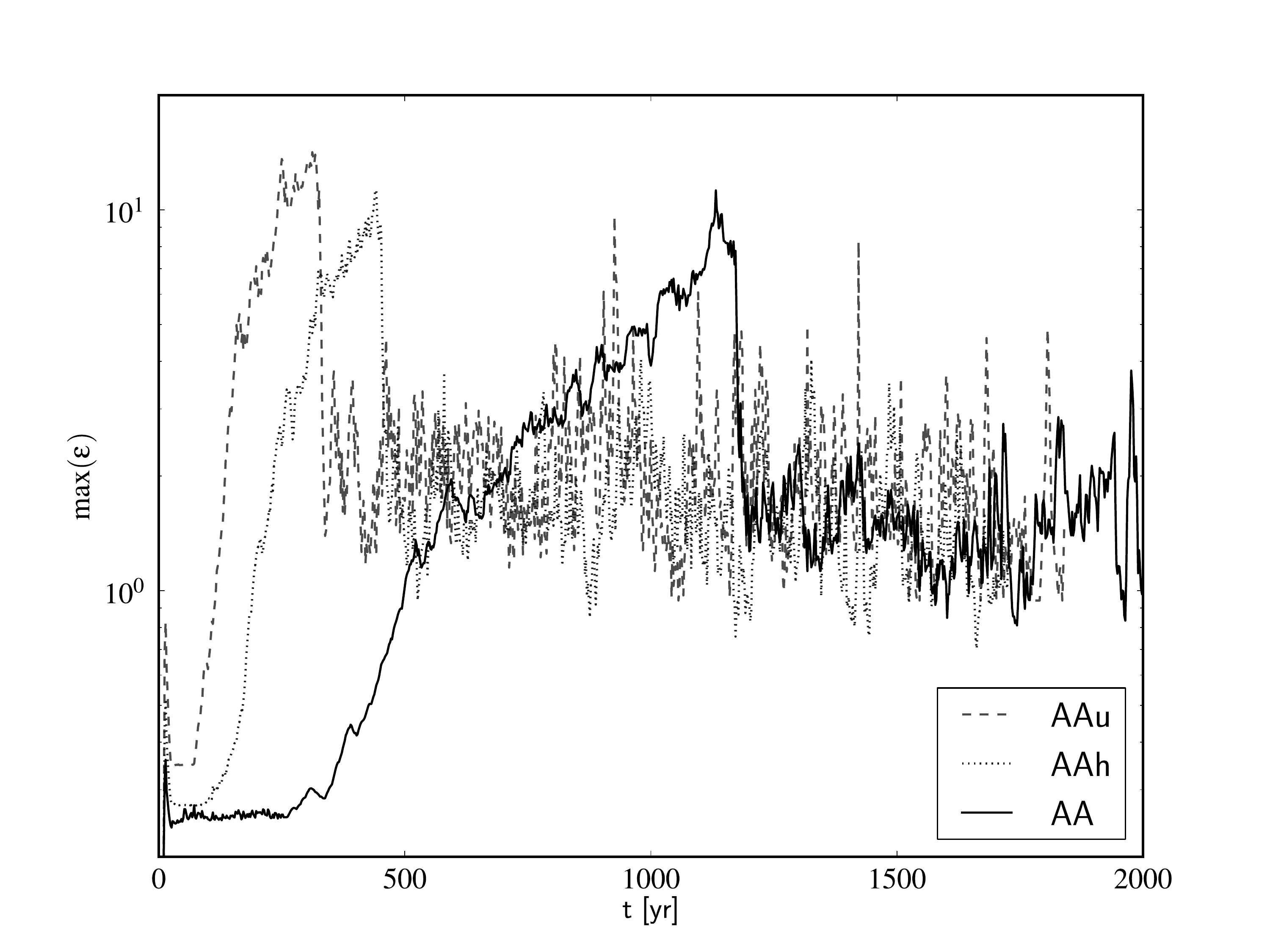}
   \caption{Maximum ratio of dust to gas density for three runs with the same
      initial conditions, but different resolution. Streaming instability
      follows regular pattern of evolution: (1) fast linear growth that locally
      increases $\epsilon$ over 2 orders of magnitude (2) after reaching certain
      level of $\epsilon \approx 10$, overdensities are abruptly smoothed out
      (3) instability reaches out saturated, non-linear phase where large,
      smooth clumps of dust are only 10 times denser than initial condition.
      Varying resolution only influences the availability of shorter and faster
      growing modes, i.e. shortens phase (1).  }
   \label{fig4}
\end{figure}

 In the gas dominated case (AA) we observe initial overdensity growth over
two orders of magnitude in a characteristic grid-like pattern.  However, at time
1200, 440, 325 for runs AA, AAh, AAu respectively, in the course of nonlinear
evolution the dense dust filaments are abruptly smoothed out resulting in
oscillatory motion of mild overdensities that are only one order of magnitude
denser than initial dust distribution (see Fig.~\ref{fig4})

\subsection{3D run}
\begin{figure}
   \includegraphics[width=0.98\linewidth]{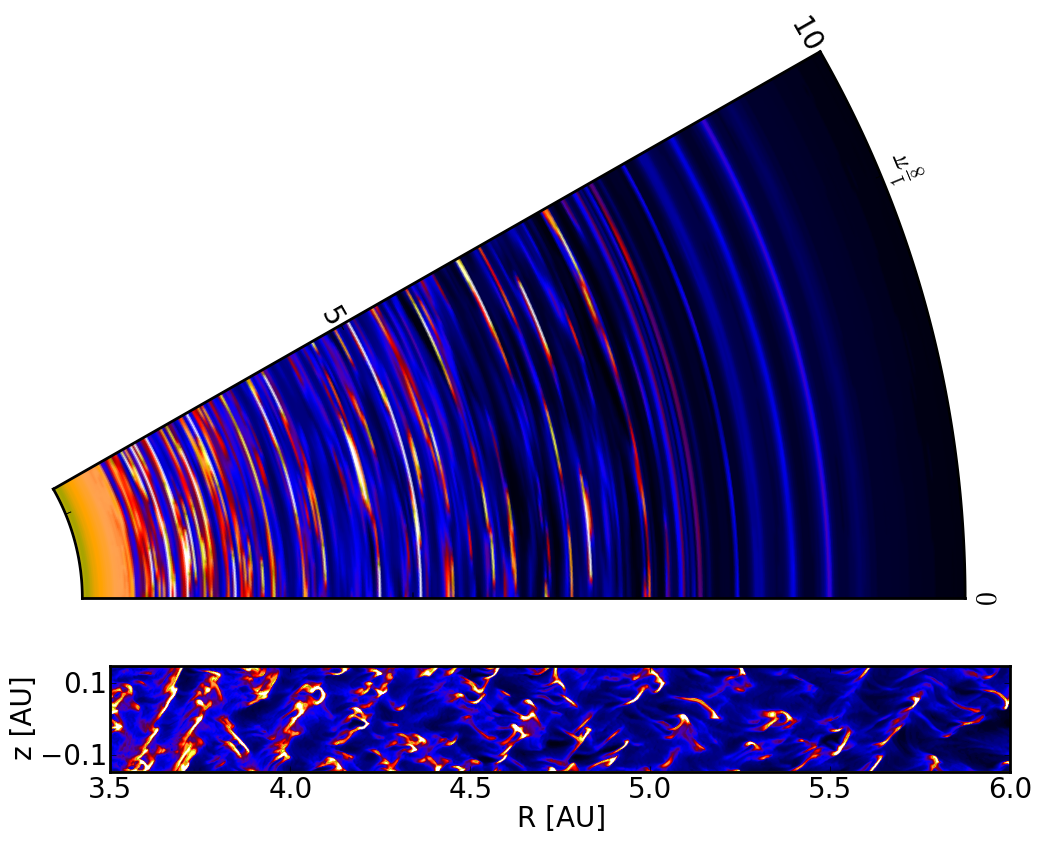}
   \caption{Snapshot of dust density distribution from run BB3D taken at
   $t=500$~yr. Upper panel shows a slice through the disc midplane, whereas lower
   panel is a part of slice perpendicular to disc plane zoom out to the region
   that corresponds to physical span of 2D simulations.}
   \label{fig5}
\end{figure}

\begin{figure} 
  \includegraphics[width=0.98\linewidth]{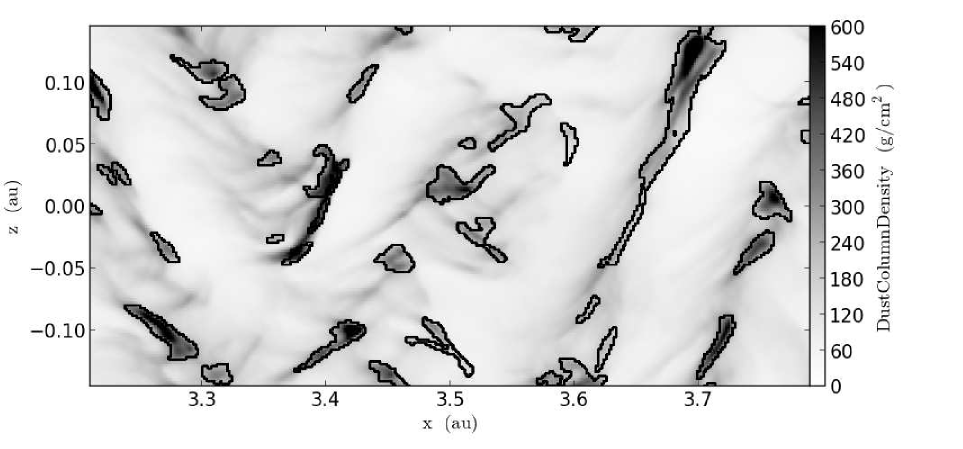}
  \caption{
     Small patch showing projection of dust density along azimuthal axis from
     run BB3D with annotated contours of gravitationally bound clumps.}
  \label{fig6} 
\end{figure}

The course of the evolution of our single 3D (BB3D) closely follows the
corresponding 2D run BB (compare Fig.~\ref{fig5} and Fig.~\ref{fig1}). During
the linear phase of growth elongated clumps or rather sheets of dense dust are
formed, as the deviation from axis axisymmetry is almost negligible.
In order to estimate whether we should expect effects of self-gravity  in the
nonlinear evolution of streaming instability, we have used \yt{}~\citep{yt} clump
finding facility to identify all the largest disconnected contour spanning for
minimum value of 50 computational cells in BB3D run. Then we calculated their
total mass and velocity dispersion $\sigma$ to obtain Jeans' radius:
\begin{equation}
   R_J = 2 GM / \sigma^2.
\end{equation}
Subsequently we compared the average size of each individual clump $L =
\sqrt{\sum_{i\in{\{x,y,z\}}} \max (L_i^2)}$ and found that $L < R_J$ for every
identified overdense clump, indicating that the clumps fulfill the gravitational
binding condition (see Fig.~\ref{fig6}). This indicates that in certain
situations, i.e. for dust dominated runs of the semi-global disc configuration,
streaming instability acting in a selfgravitating medium can lead to planetesimals
formation~\citep{J07}.

\subsection{Comparison to the Results of Linear Stability Analysis
\label{simulation_analysis}}

In order to analyse the growth of streaming instability we extract small square
patches from various locations across the computational domain. The patches have
sizes of $0.15^2$~AU$^2$, thus are small enough so that mean values of physical
quantities do not vary significantly across them. In the case of 3D run the
patches are chosen on the {\it r-z} plane at $\varphi = \varphi_\textrm{max} /
2$. 

We note that in the place of the fixed parameters  describing the unperturbed
equilibrium  in \mref{lin1}--\mref{lin2}, we use patch-averaged values of
corresponding dependent variables in equations \mref{eqc1}--\mref{eqm2}. 
We calculate the spatially averaged densities of gas and dust component
$\bar{\rho}_g = \left<\rho_g\right>$, $\bar{\rho}_d = \left<\rho_d\right>$ and
their mutual ratio $\bar{\epsilon} = \left<\rho_d / \rho_g\right>$ . Mean
angular velocity $\bar{\Omega}$ is taken for the center of the patch.
Dimensionless measure of sub-Keplerian gas rotation is calculated by means of
the formula~(see YG05 eq.~(16) or JY07 eq.~(1))
\begin{equation}
   \bar{\eta} = -\frac{c_s^2\left<\partial_R \left<\rho_g\right>_z\right>_R}
      {2\bar{\rho}_g\bar{\Omega}^2 R},
   \label{eq:eta}
\end{equation}
In formula \mref{eq:eta} we average gas density in the vertical direction, then
we calculate the mean radial derivative of $\left<\partial_R \rho_g\right>$. 
The expression for mean stopping time is derived from \mref{eq:tauf}
\begin{equation}
   \bar{\tau}_f = \rho_\bullet a / \left(\bar{\rho}_g \sqrt{c_s^2 +
   \left<\left|\mathbf{u} - \mathbf{v}\right|^2\right>} \right).
\end{equation}
For the sake of consistency gas and dust velocities $\bar{\mathbf{u}},
\bar{\mathbf{w}}$ are also approximated by their mean values
$\left<\mathbf{u}\right>, \left<\mathbf{w}\right>$ instead of calculating them
with the aid of \mref{eq:w0}-\mref{eq:u0}. However, we note that the mean values
that are naturally achieved during the simulation,  do not diverge from the
equilibrium solution by more than $10\%$. 

\begin{figure}
  \includegraphics[width=0.98\linewidth]{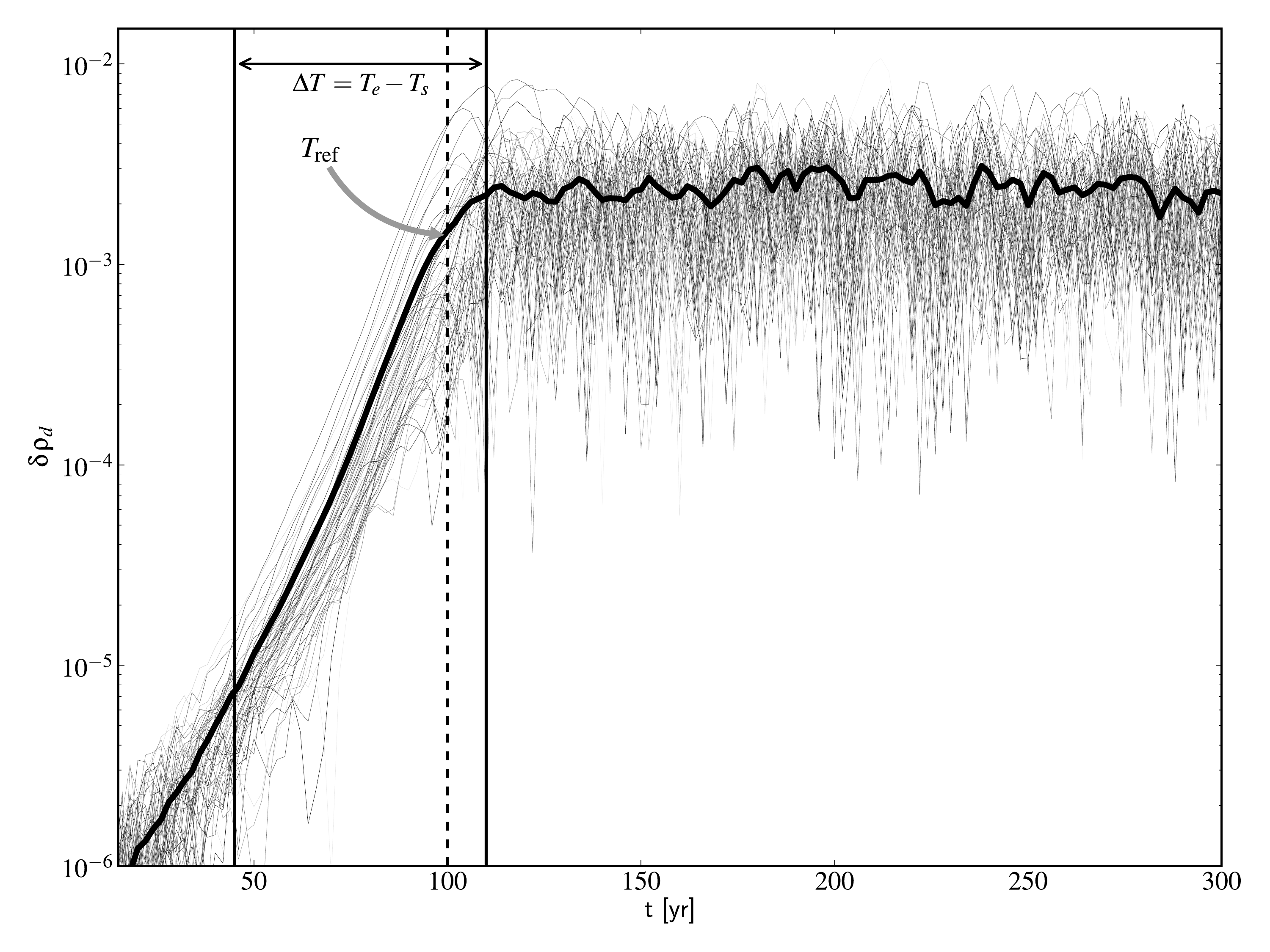}

  \caption{Temporal evolution of dust density perturbation amplitude in the
     Fourier space in the patch located at $[2.85,3.15]\times[-0.15,0.15]$~AU
     from simulation BB.  Each line represents amplitude for a given pair of
     $k_x, k_z$.  $\Delta T = T_e - T_s$ is time range when modes are fitted
     with \mref{eq:fit}.  $T_{\textrm{ref}}$ is a reference point in time used
     to identify most dominant modes using criterion of maximum amplitude value.
     Thick line is an average of amplitudes that are grater than $10^{-4}$ at
     reference time $t = T_{\textrm{ref}}$
   } 
   \label{fig7} 
\end{figure}

To determine linear growth rates of the excited instability modes we calculate
Fourier transforms of density and velocity distributions of the dust component.
We analyse subsequently time variation of amplitudes of individual modes.
Intermediate results are shown in Fig.~\ref{fig7}. 

We identify time $T_s$ at which the dominating modes emerge from fluctuations
and start their linear growth phase. Similarly, we identify the end of the
linear growth phase $T_e$ when modes have grown by few orders of magnitude, and
start to saturate their growth. We fit exponential function:
\begin{equation}
   f(t) = A\exp\left(-s t\right)
   \label{eq:fit}
\end{equation}
 to the measured density amplitudes in the  period $\Delta T = T_e - T_s$.

That procedure allows us to determine the growth rate $s(k_x, k_z)$ of each
individual unstable mode, during the linear phase of the instability growth in
the numerical experiments.  We identify the most rapidly growing modes by
selecting those with the greatest amplitude at fixed point at the reference time
$T_{\textrm{ref}}$ before the saturation. We compare the growth rates $s(k_x,
k_z)$ obtained for each mode  with the solution $s_0(k_x, k_z)$ resulting from
the linear analysis for the local mean flow parameters. 

\begin{figure} 
   \includegraphics[width=0.98\linewidth]{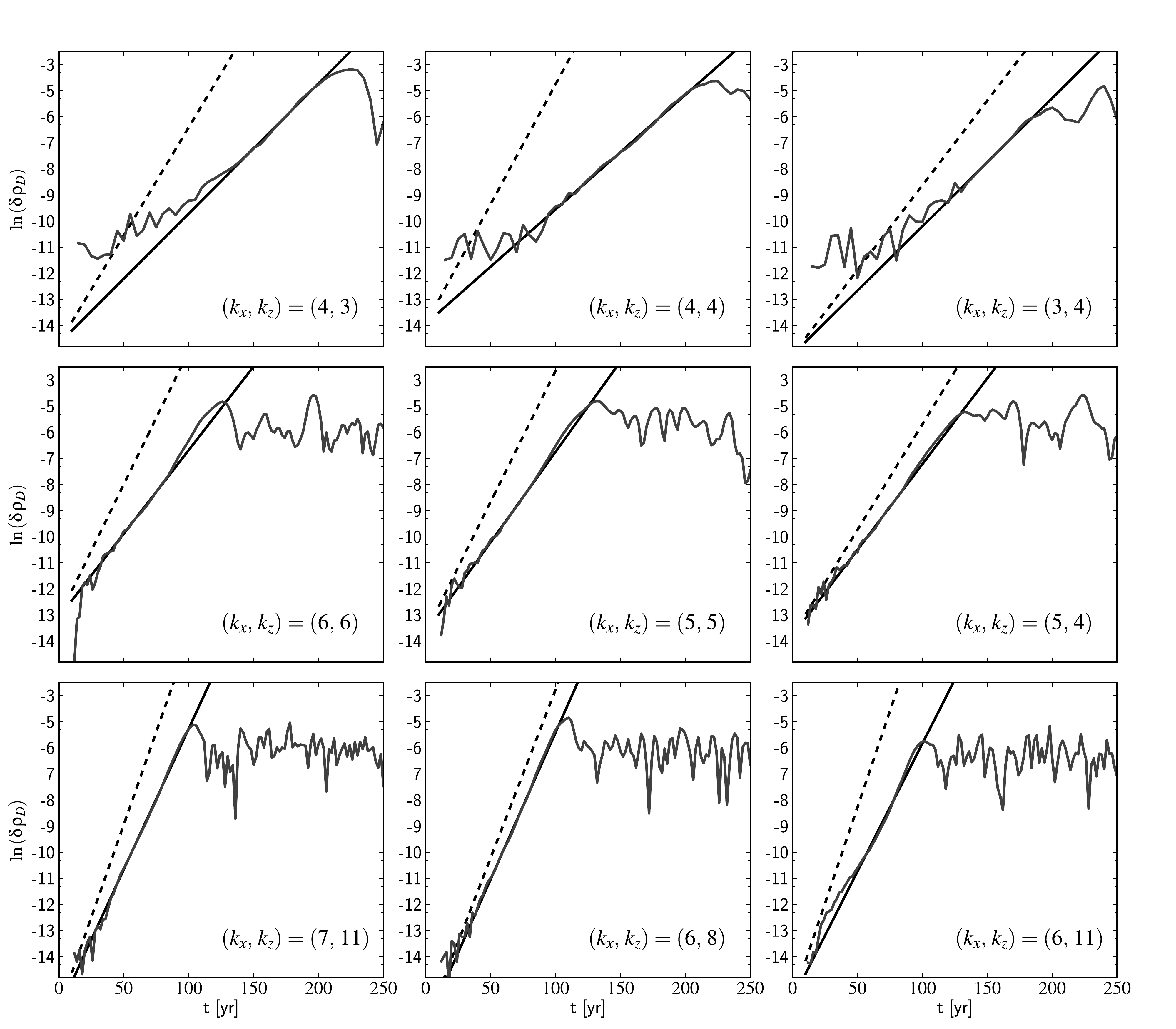}
   \caption{Temporal evolution of the amplitudes of the dominating instability
      modes, measured in the dust density (gray line) together with the lines
      resulting from the linear analysis of the streaming instability (dashed
      line) and lines of fits of eq.~\mref{eq:fit} to the measured amplitudes
      (black lines).
      The linear solution of eq.~\mref{eq:linset} is found for the mean
      parameters of the rectangular patch at $R=3$~AU for the simulations: BB3d
      (upper panel), BB (middle panel) and BBh (lower panel). 
   }
   \label{fig8}
\end{figure}

In Fig.~\ref{fig8} we show temporal evolution of dust density perturbation
amplitudes for three dominating instability modes in three experiments BB3d, BB
and BBh, together with lines fitted to the phase of linear growth and  lines
representing growth of the amplitudes predicted by the linear analysis of the
streaming instability.  In the mid resolution run BB the growth rates  are
$10\div30\%$ smaller than those predicted by linear analysis, what indicates
that still higher resolution is required to resolve these modes. In the high
resolution run BBh the results of the numerical experiment tend to converge the
results of linear stability analysis.  The final saturation amplitudes seem to
be slightly smaller in high resolution runs. One should note, however, that we
have chosen the modes of the highest growth rate for each run, which are not the
same modes in terms of wavenumbers. 
 
\par
\begin{figure*}
  \includegraphics[width=0.48\linewidth]{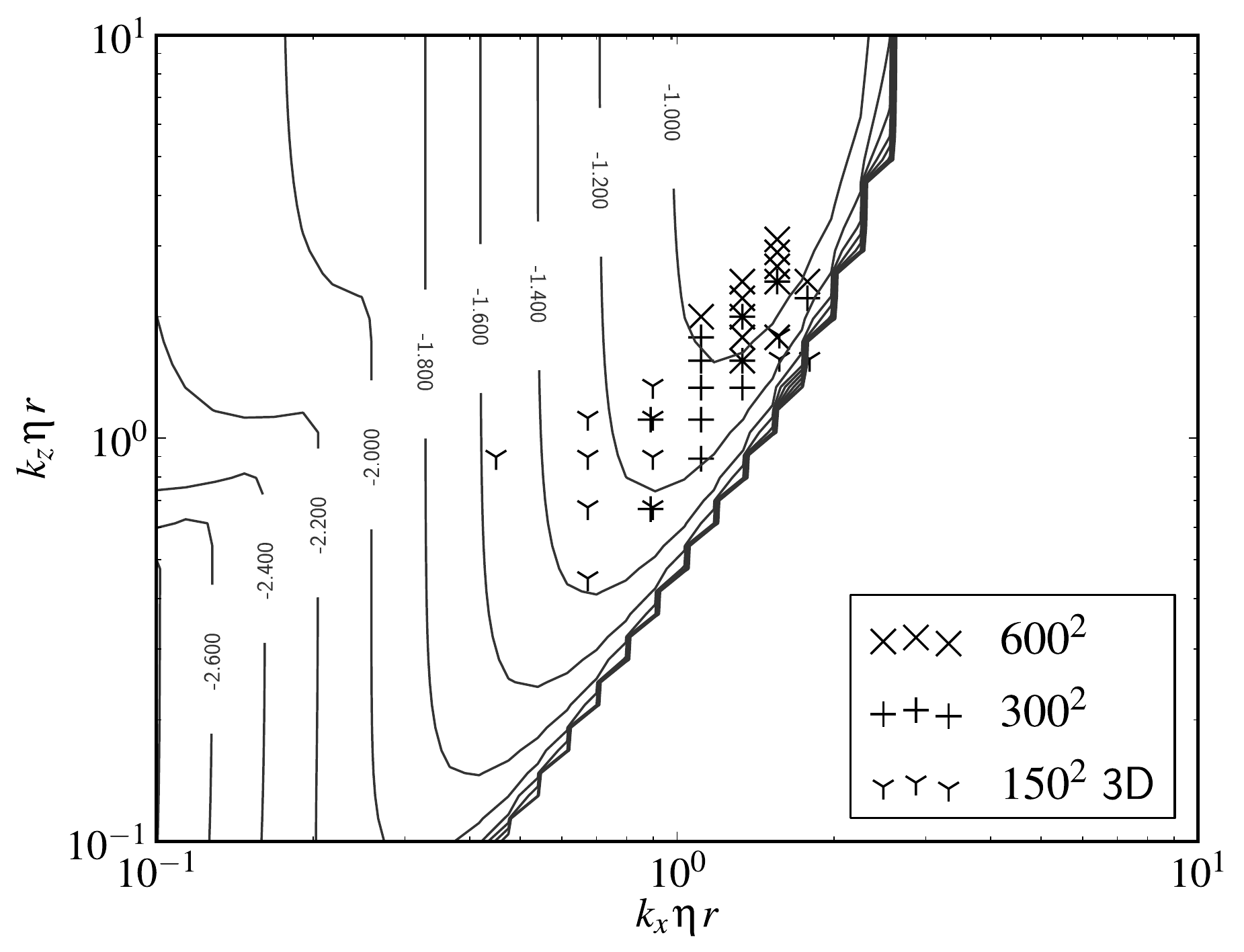}
  \includegraphics[width=0.48\linewidth]{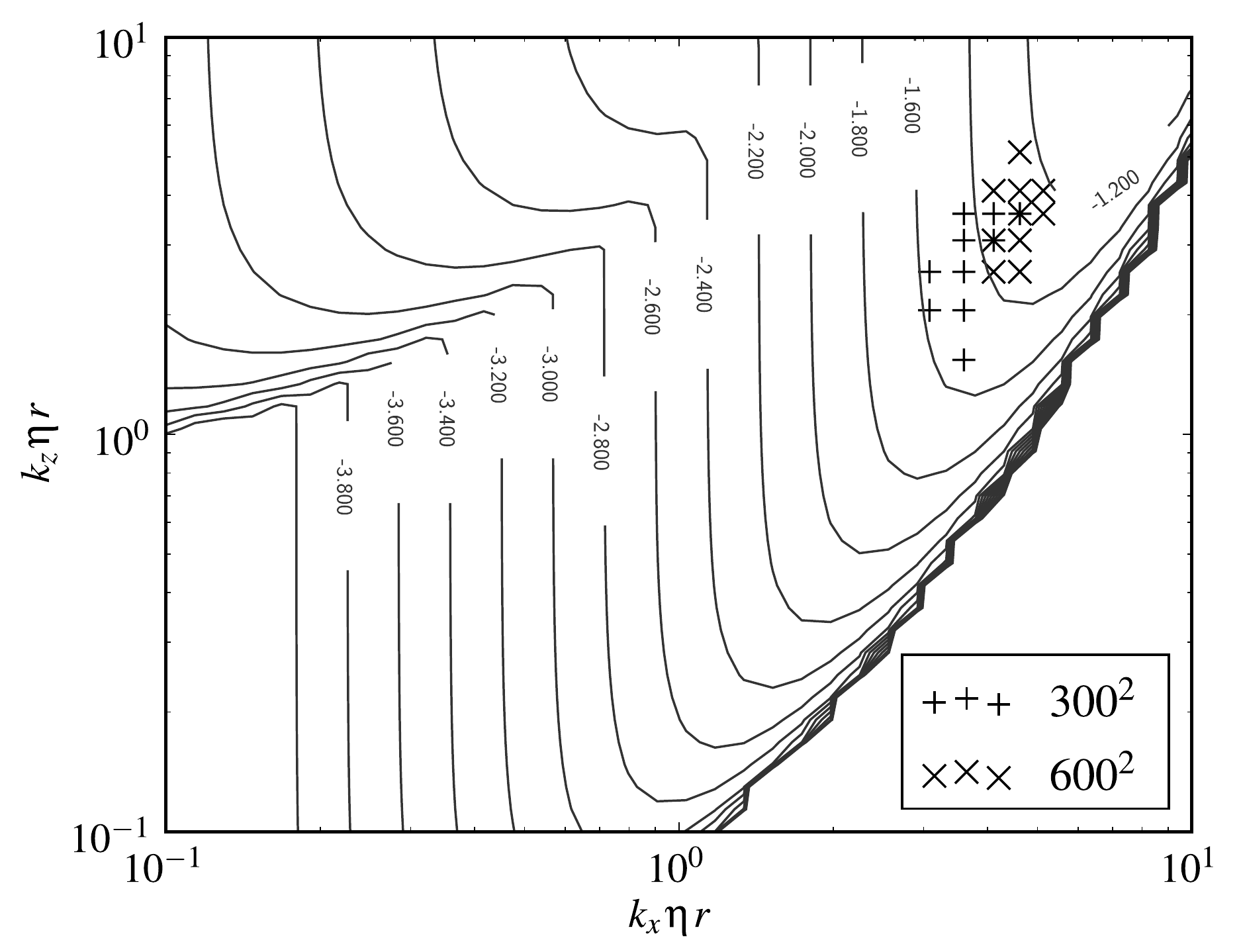}
  \caption{
     The nine of the fastest growing modes characterized by wavenumbers
     $(k_x, k_z)$ extracted  from different simulations sharing the
     same initial conditions (left panel: BB, BBh, BB3d, right panel: AB, ABh)
     but different resolutions. Contours of growth rates $\log_{10}( s(k_x,
     k_z))$  represent solutions of \mref{eq:disprel} for the mean state
     parameters derived for  the domain patches at $T = T_{\textrm{ref}}$
  (for comparison see Fig.~2 of YG05)}
   \label{fig9}
\end{figure*}

 To extend our analysis we show in Fig.~\ref{fig9} a contour plot
(following YG05 and YJ07) of the growth rate  $s(k_x, k_z)$, resulting from
solutions of the dispersion relation \mref{eq:disprel}.  The shape of
isocontours indicates that the most unstable modes of the streaming instability,
predicted by linear analysis, form a ridge extending towards  large values of
$k_x$ and $k_z$.  The plots are constructed for the same set of the physical
parameters and the mean state parameters derived for  the domain patches at $T=T
_{\textrm{ref}}$.  We then place  $(k_x, k_z)$ of 9 dominating modes extracted
at $T = T_{\textrm{ref}}$  for simulations performed at different resolutions.

Our procedure allows us to confirm that  wavenumbers  of the modes, emerging in
presented numerical models, align with the  ridge of linear solutions at the
growth rate map.  It is apparent that dominating modes  are limited by available
numerical resolution. At lower numerical resolutions (run BB3d)  the dominant
modes locate below the contour labeled as ''$-1.000$''. At the mid resolution of
run BB some of the modes locate above the contour ''$-1.000$'', and at the high
resolution most of the modes locate above this contour. Similar tendency can be
observed for the pair of runs AB and ABh.  We anticipate that the reason for the
absence of the very short wavenumber modes is the numerical diffusivity of the
presently  used Relaxing TVD algorithm. Our estimations show that  our code
requires at least 32 computational cells per wavelength of the unstable mode, to
accurately represent the linear growth rate, which is much more than is required
by higher-order schemes~\cite{YJ07, BT09}.

\begin{figure}
   \includegraphics[width=0.98\linewidth]{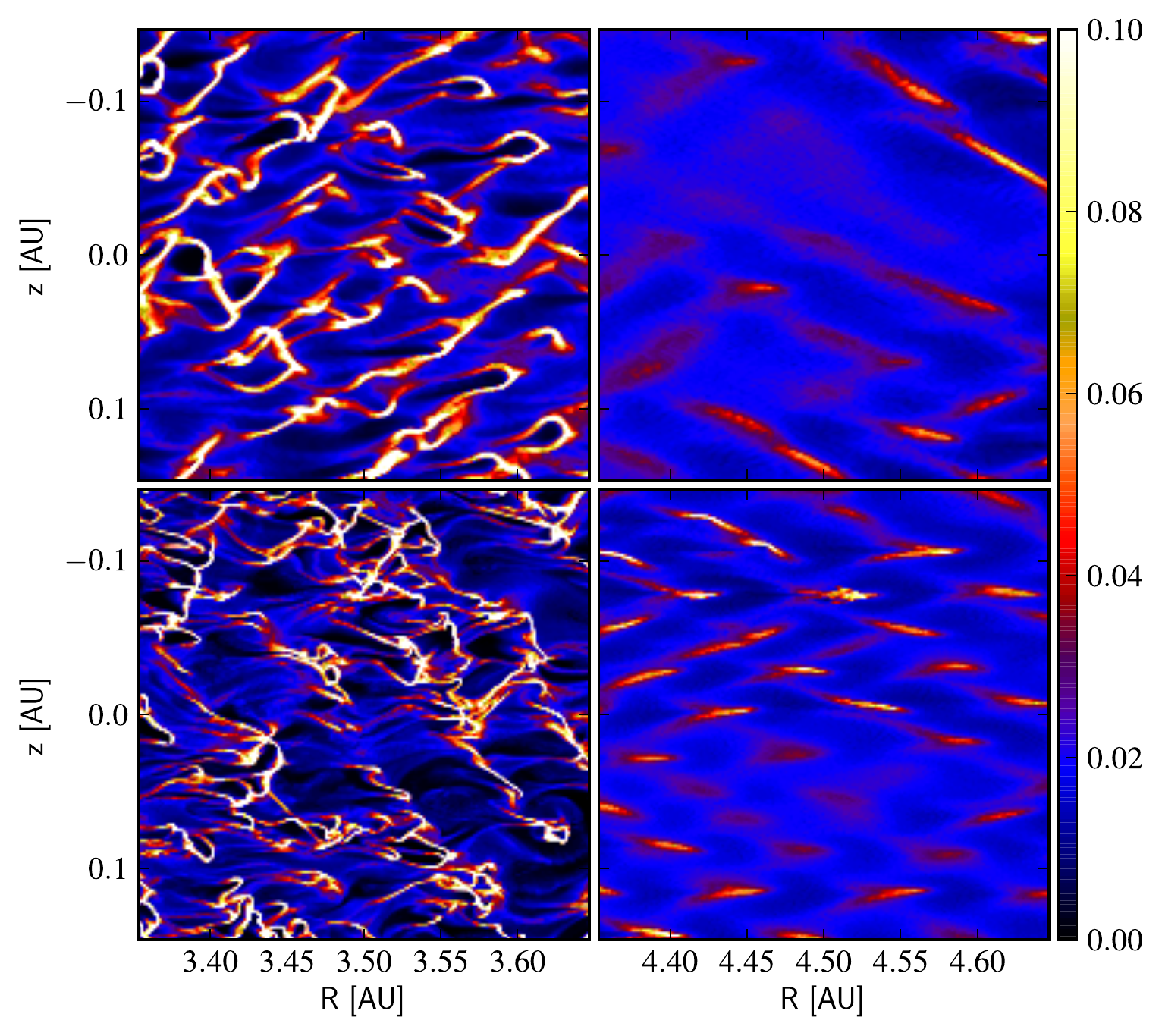}
   \caption{Dust density snapshots at $t = 160$~yr in two small rectangular
      patches for the two identical simulations parameters (upper: BB, lower:
      BBh) showing how the resolution affects simulation results at the linear
      and saturation phase of the instability. It is apparent that shorter
      wavelength modes dominate at the higher resolution, and that dust
      condensation structures are sharper.
   } 
   \label{fig10} 
\end{figure}

We varied the resolution of our simulations to check how our numerical scheme
affects the obtained solutions. As the streaming instability in general
"prefers" shorter wavelengths for the optimal growth, increasing resolution
always leads to more, smaller overdensities emerging during the course of
evolution (see Fig.~\ref{fig10}). However we note that our results
follow the fastest growing, linear modes (see Fig.~\ref{fig9}) and
that most phenomena described in previous sections i.e. cavitation (see
Fig.~\ref{fig3}) or sudden growth dumping of the tightly coupled
boulders in the gas dominated regime (see Fig.~\ref{fig4}) are independent of
the size of the smallest computational cell.

\section{Discussion \& Conclusions}
\label{sec:conclusions}
We performed a set of 2D simulations and 3D simulation of two mutually
interacting fluids, i.e. gas and dust in protoplanetary disc, while neglecting
vertical component of gravity. For the spectrum of grain sizes ranging from
$10\div50$~cm we observe rapid growth of unstable modes in dust component, which
are consistent with modes obtained for streaming instability within the
framework of local linear analysis.

Our model enhances previous work of other authors by taking into account the
full dynamics of protoplanetary disc, e.g. radial migration, that lead to
significant variation in physical quantities, such as gas pressure gradient, and
were previously treated as constant. Moreover instead of using a general
dimensionless stopping time parameter to couple both fluid components, we
implement specific drag law (Eqn.~\mref{eq:tauf}) which is best suited to the
physical properties of simulated fluids.

\par In the present quasi-global approach we relieved the restriction of the
fixed  radial pressure gradient, adopted for numerical modeling of the streaming
instability within the shearing box approximation.  It remains unknown how much
this term affects the non-linear evolution of the system, since the additional
term acts as an infinite energy reservoir driving the difference in azimuthal
velocities of both fluids. In the global approach, even around fixed orbit, the
parameter $\eta$, measuring the radial pressure gradient, is function of time.
While locality of shearing box fully justifies using a constant dimensionless
stopping time to calculate mutual linear drag force, that approximation is no
longer valid in the global disc. 

\par The wave numbers of the fastest growing modes, extracted from numerical
models, and their positions at  the  map of growth rate  $s$ vs. $k_x$ and $k_z$
coincide with the predictions of linear analysis of streaming instability. We
reproduce reasonably well the linear growth rates, getting rapid convergence
near $\sim 32$ cells per wavelength. It is apparent, however,  that due to
diffusive nature of the Relaxing TVD scheme, PIERNIK needs high resolution to
capture linear growth phase accurately.

\par The instability saturates when dust overdensities synchronize their
velocity with the gas component. We must note that despite the steady outward
migration of the gas, dust blobs never really gain positive radial velocity and
never cease to migrate inwards (this was also noted by~\cite{JY07}). Though in
certain cases, their radial velocity is significantly slower than what is
expected from the standard estimation of the migration rate. The lack of
additional force that would keep the blobs bounded also leads to high dispersion
rate of the overdensities.

\par In our simulations we have neglected the role of stratification of the
disc, self-gravity of both fluids and also MRI induced turbulence. As each and
every of the aforementioned physical processes plays significant and integral
role in the protoplanetary disc evolution, we plan to gradually expand our setup
increasing its complexity in order to create the more complete global model.

\par We have found that nonlinear evolution of  the streaming instability forms
conditions for the formation  of gravitationally bound dust blobs in the
examined 3D semi-global discs configuration.  This indicates, in accordance with
the predictions by~\citet{J07}, the possibility  for  planetesimals formation in
accreting circumstellar discs,  and therefore we plan to incorporate selfgravity
in our future work.

\par We note that some of our runs may be underresolved, especially the
gas-dominated setups. However the nonlinear outcomes of all setups converge to
common solutions. Moreover, our initial simulation of full 3D setup, which took
0.36 MCPUh, shows that we are able to perform global simulation of streaming
instability in a convincing manner. 
However, due to the severe CFL constraint on time-step imposed by the high
azimuthal velocity in 3d simulations, we consider implementing {\sc FARGO}-like
algorithm~\cite{M00} for our future work.
This is an important step towards complete model of planetesimals formation due
to combined action of various fluid instabilities postulated by~\citet{J07}. 
\section*{Nullius in verba}
We would like to stress that everything that is required to reproduce results
presented in this paper is publicly available: the code
(\url{http://piernik.astri.umk.pl}), initial conditions (part of PIERNIK's
example problem set -- {\it streaming global}), analysis and visualization
routines (aforementioned github repository and \yt{} package). Runs in the
basic resolution took approx. 400 CPUh and can be conveniently reproduced on
small clusters or even modern workstations. KK is willing to assist inquisitive
readers with performing said simulations along with providing access to data
obtained in much more computationally demanding runs.
\section*{Acknowledgments}

This work was partially supported by Polish Ministry of Science and Higher
Education through the grant N203 511038. This work is a part of POWIEW project
supported by the European Regional Development Fund in the Innovative Economy
Programme (POIG.02.03.00-00-018/08). This research was supported in part by
PL-Grid Infrastructure. All of the post-processing and procedures described in
Sect. \ref{simulation_analysis}  was carried out using the data analysis and
visualization package \yt{} \footnote{\url{http://yt-project.org/}}
by~\cite{yt}.

\label{lastpage}

\begin{thebibliography}{99}
\bibitem[\protect\citeauthoryear{Bai \& Stone}{2010a}]{BS10a} 
 Bai X.-N., Stone J.~M., 2010, ApJ,
 722: 1437
\bibitem[\protect\citeauthoryear{Bai \& Stone}{2010b}]{BS10b}
 Bai X.-N., Stone J.~M., 2010, ApJ,
 722: L220
\bibitem[\protect\citeauthoryear{Balbus \& Hawley}{1998}]{BH98}
 Balbus S.~A., Hawley J.~F. 1998, RvMP,
 70: 1
\bibitem[\protect\citeauthoryear{Balsara et al.}{2009}]{BT09}
 Balsara, D.~S., Tilley, D.~A., Rettig, T., Brittain, S.~D., 2009, MNRAS
 397: 24
\bibitem[\protect\citeauthoryear{Blum \& Wurm}{2008}]{BW08}
 Blum J., Wurm G., 2008, ARA\&A, 
 46: 21
\bibitem[\protect\citeauthoryear{Cuzzi et al.}{1993}]{CD93}
 Cuzzi J.~N., Dobrovolskis A.~R., Champney J.~M., 1993, Icarus
 106: 102
\bibitem[\protect\citeauthoryear{Durand}{1960}]{D60}
 Durand, E., 1960,
 In Masson et al. {\it Solutions Numériques des Equations Algébriques},
 vol. 1, pp. 279--281 
\bibitem[\protect\citeauthoryear{Kerner}{1966}]{K66}
 Kerner, I.~O., 1966, Numerische Mathematik 
 8: 290-294.
\bibitem[\protect\citeauthoryear{Goldreich \& Ward}{1973}]{GW73}
 Goldreich P., Ward W.~R., 1973, ApJ,
 183: 1051
\bibitem[\protect\citeauthoryear{Hanasz et al.}{2010}]{piernik2} 
 Hanasz M., Kowalik K., W\'olta\'nski D., Paw\l{}aszek, R.~K., Kornet, K., 2010,
 {\it EAS Publications Series}
 42, 281
\bibitem[\protect\citeauthoryear{Hayashi}{1981}]{H81} 
 Hayashi, C., 1981, Prog. Theor. Phys. Suppl., 
 70, 35
\bibitem[\protect\citeauthoryear{Hawley et al.}{1995}]{HGB95}
 Hawley J.~F., Gammie C.~F., Balbus S.~A., 1995, ApJ, 
 440, 742
\bibitem[\protect\citeauthoryear{Kokubo et al.}{2006}]{KKI06}
 Kokubo E., Kominami J., Ida S., 2006, ApJ, 
 642, 1131
\bibitem[\protect\citeauthoryear{Kosi\'nski \& Hanasz}{2006}]{KH06}
 Kosi\'nski, R., Hanasz, M., 2006, MNRAS,
 368, 759
\bibitem[\protect\citeauthoryear{Jacquet et al.}{2011}]{J11}
 Jacquet E., Balbus S., Latter H., 2011, MNRAS,
 415, 3591
\bibitem[\protect\citeauthoryear{Johansen et al.}{2006}]{JHK06}
 Johansen A., Henning T., Klahr H., 2006, ApJ,
 643, 1219
\bibitem[\protect\citeauthoryear{Johansen et al.}{2007}]{J07}
 Johansen A., Oishi J.~S., Mac Low M.-M., Klahr H., Henning T., Youdin A.,
 2007, Nature, 448, 1022
\bibitem[\protect\citeauthoryear{Johansen \& Youdin}{2007}]{JY07}
 Johansen A., Youdin A., 2007, ApJ,
 662, 627
\bibitem[\protect\citeauthoryear{Lee et al.}{2010}]{L10}
 Lee A.~T., Chiang E., Asay-Davis X., Barranco J., 2010, ApJ,
 725, 1938
\bibitem[\protect\citeauthoryear{Lesur \& Papaloizou J.~C.~B.}{2010}]{LP10}
 Lesur G., Papaloizou J.~C.~B., 2010, AA,
 513, 60
\bibitem[\protect\citeauthoryear{Masset}{2000}]{M00}
 Masset F., 2000, A\&AS,
 141, 165
\bibitem[\protect\citeauthoryear{Mignone et al.}{2007}]{M07}
 Mignone, A., Bodo, G., Massaglia, S., Matsakos, T., Tesileanu, O., Zanni, C.,
 Ferrari, A., 2007, ApJS,
 170, 228
\bibitem[\protect\citeauthoryear{Nakagawa et al.}{1986}]{N86}
 Nakagawa Y., Sekiya M., Hayashi C., 1986, Icarus,
 67, 375
\bibitem[\protect\citeauthoryear{Perucho et al.}{2004}]{PHM04}
 Perucho, M., Hanasz, M., Mart{\'{\i}}, J.~M., Sol, H., 2004, A\&A,
 427, 415
\bibitem[\protect\citeauthoryear{Shakura \& Sunyaev}{1973}]{SS73}
 Shakura N.~I., Sunyaev R.~A., 1973, A\&A, 
 24, 337
\bibitem[\protect\citeauthoryear{Skinner \& Ostriker}{2010}]{SO10}
 Skinner, M.~A., Ostriker, E.~C., 2010, ApJSS 
 188, 290
\bibitem[\protect\citeauthoryear{Takeuchi et al.}{2012}]{T12} 
 Takeuchi T., Muto T., Okuzumi S., Ishitsu N., Ida S., 2012, ApJ,
 744, 101
\bibitem[\protect\citeauthoryear{Tilley et al.}{2009}]{TB09}
 Tilley D.~A., Balsara D.~S., Brittain S.~D., Rettig T., 2009, MNRAS,
 403, 211
\bibitem[\protect\citeauthoryear{Turk et al.}{2011}]{yt}
 Turk, M.~J., Smith, B.~D., Oishi, J.~S., Skory, S. Skillman, S.~W., Abel, T.,
 Norman, M.~L., 2011, ApJS,
 192, 9
\bibitem[\protect\citeauthoryear{Weidenschilling}{1977}]{W77} 
 Weidenschilling S.~J., 1977, MNRAS,
 180, 57
\bibitem[\protect\citeauthoryear{Youdin \& Goodman}{2005}]{YG05} 
 Youdin A., Goodman, J., 2005, ApJ,
 620, 459
\bibitem[\protect\citeauthoryear{Youdin \& Johansen}{2007}]{YJ07}
 Youdin A., Johansen A., 2007, ApJ,
 662, 613
\bibitem[\protect\citeauthoryear{Zsom et al.}{2010}]{Z10} 
 Zsom A., Ormel C.~W., G\:uttler C., Blum J., Dullemond C.~P., 2010, AA,
 513, 57
%
\end{thebibliography}
\end{document}